# Improvement to the Prediction of Fuel Cost Distributions Using ARIMA Model

Zhongyang Zhao, *Student Member, IEEE,* Chang Fu, *Student Member, IEEE*, Caisheng Wang, *Senior Member, IEEE*
Department of Electrical and Computer Engineering, Wayne State University, Detroit, USA
Carol J. Miller
Department of Civil and Environmental Engineering, Wayne State University, Detroit, USA

*Abstract*— Availability of a validated, realistic fuel cost model is a prerequisite to the development and validation of new optimization methods and control tools. This paper uses an autoregressive integrated moving average (ARIMA) model with historical fuel cost data in development of a three-step-ahead fuel cost distribution prediction. First, the data features of Form EIA-923 are explored and the natural gas fuel costs of Texas generating facilities are used to develop and validate the forecasting algorithm for the Texas example. Furthermore, the spot price associated with the natural gas hub in Texas is utilized to enhance the fuel cost prediction. The forecasted data is fit to a normal distribution and the Kullback-Leibler divergence is employed to evaluate the difference between the real fuel cost distributions and the estimated distributions. The comparative evaluation suggests the proposed forecasting algorithm is effective in general and is worth pursuing further.

## I. INTRODUCTION

Various smart grid technologies have been proposed to modernize the existing power grid. However, prior to application, the proposed methods must be validated with test data and/or historical conditions using validated, realistic, high fidelity power system network models. Important input variables to the development of such models include fuel costs (FCs), heat rates, generation costs and emission rates associated with each class of prime mover and fuel type combination. Extensive research has been carried out to provide related information [1], [2]. The US Energy Information Administration (EIA)-923 form collects detailed monthly and annual electric power data and can be used in developing fuel cost model. In [3], a FC characterization and distribution method is carried out based on publicly available data. The EIA-923 Form [4] is used in [3] to calculate the FC distributions for different area-fuel type combinations based on different plant IDs, fuel quantities, average heat contents, and so on. The FC distributions can be used to generate large and realistic power system network models without compromising the confidentiality of the utilities. However, the EIA-923 form updates the fuel cost data with a three-month delay. The information available is the data three months before. Due to the three months update lagging of the form, the most current data is not available for the study. Hence, the motivation of this paper is to develop a forecasting algorithm to provide more accurate FC characterization and distribution estimation instead of relying on the data with a three-month delay. The refined fuel cost distribution produced by the proposed method in this paper will be able to better serve for a marginal emission estimation model called Locational Emission Estimation Methodology (LEEM) [5]. LEEM automatically tracks, analyzes and reports location-specific, real-time and forecasted marginal emissions information for empowers utility services, energy users, data and energy management firms, government, and individuals to predict emission levels and estimate costs over time and make better, more informed energy and emission management decisions for the future.

As an effective time series analysis tool, an ARIMA model is implemented to forecast price signals, such as fuel costs and natural gas hub spot prices in this paper. ARIMA models have been analyzed and evaluated for forecasting natural gas price and electricity price [6]-[8]. The ARIMA models using standard errors based on Hessian, Cochrane-Orcutt, Prais-Winstern and Hildreth-Lu have been developed to forecast the price of natural gas in [6]. For forecasting the Henry Hub weekly natural gas spot price, the ARIMA model aiming for the approximation components associated with wavelet decomposition was implemented [7]. Meanwhile, ARIMA model is also popular to forecast other dynamic series, such as electricity price. The differential electricity price between the day-ahead locational marginal price and the real-time locational marginal price was derived and forecasted by ARIMA models [8]. The similar differential series implemented on ARIMA is developed in this paper as well.

This paper first explores the monthly data features of Form EIA-923 from January 2013 to December 2016 and determines the natural gas (NG) fuel cost of Texas as the objective target to predict and estimate distributions to validate the forecasting method proposed in this paper. The differential series between the plant's fuel cost and NG hub spot price is formulated and implemented on the ARIMA model for a three-step-ahead prediction. As an enhancement to the fuel cost prediction, the corresponding local natural gas hub spot price is utilized and predicted by one-step-ahead. In Section IV, the fuel cost data between January 2013 and June 2016 are used to train the formulated ARIMA model. The next six months data are adopted to validate the proposed forecasting algorithm. The results show the proposed method is able to achieve outstanding performance when the overall market price is not volatile. Besides, this method shows the capability to forecast the price after the price turning point appears.

## II. DATA FEATURES

In this paper, the public fuel cost data extracted from Form EIA-923 [4] are analyzed and utilized for achieving more accurate probability density estimation using ARIMA model forecasting.

## A. Form EIA-923 Exploration

According to Form EIA-923, covering the monthly fuel cost data from 2013 January to 2016 December, the Date (Year and Month), Plant ID, Plant State, Energy Source, Quantity, Average Heat Content and Fuel Cost are archived and shown in Table I for reference.

Table I  RAW DATA IN FORM EIA-923

| Date | Plant ID | Plant State | Energy Source | Q (ton) | AHC (MMBtu/ton) | FC ($/MMBtu) |
|---|---|---|---|---|---|---|
| 201301 | 127 | TX | SUB | 476 | 17.7 | 1.10 |
| 201301 | 127 | TX | SUB | 28866 | 17 | 2.29 |
| 201301 | 127 | TX | SUB | 43824 | 16.8 | 2.08 |
| 201301 | 127 | TX | SUB | 86254 | 17 | 2.12 |
| 201301 | 127 | TX | SUB | 43256 | 17 | 2.10 |

As shown in Table I, multiple fuel cost data information for same Energy Source of a plant in a single month can be found. Hence, the fuel cost data are required to be grouped together by Date, Plant ID and Energy Source for obtaining an updated overall fuel cost for each Energy Source in a month. The updated fuel cost $FC'$ are calculated by (1), (2) and (3):

$$Total\ Cost_{\substack{Plant\ ID=i,\\Date=ym,\\Source=type}} = \sum_{\substack{Plant\ ID=i,\\Date=ym,\\Source=type}} Q \times AHC \times FC \quad (1)$$

$$Total\ Heat_{\substack{Plant\ ID=i,\\Date=ym,\\Source=type}} = \sum_{\substack{Plant\ ID=i,\\Date=ym,\\Source=type}} Q \times AHC \quad (2)$$

$$FC'_{Plant\ ID,Date,Source} = \frac{Total\ Cost_{Plant\ ID,Date,Source}}{Total\ Heat_{Plant\ ID,Date,Source}} \quad (3)$$

where $Total\ Cost_{Plant\ ID=i,Date=ym,Source=type}$ represents the total fuel cost calculated by the Quantity ($Q$), the Average Heat Content ($AHC$), and the Fuel Cost ($FC$) of the plant with ID = $i$ (e.g., 127) when Data = ym (e.g., 201301) for Energy Source = type (e.g., SUB). Similarly, $Total\ Heat$ stands for the total heat of the corresponding plant with respect to the specified date and energy source.

## B. Objective Data Selection

While the gross Quantity is aggregated to each state in Form EIA-923 from 2013 to 2016, Texas is found to be the top state in the energy consumption sources and is taken as an example for the study in this paper. Meanwhile, by combining the quantity of each energy source, natural gas (NG), Bituminous coal (BIT), Sub-bituminous coal (SUB) are discovered to be the top 3 consumed energy sources from 2013 to 2016. Hence, the NG, SUB and BIT in Texas are aimed to be analyzed subsequently.

After the $FC'$ are updated and associated Plant ID, Energy Source and Date by (1), (2) and (3), the NG, SUB and BIT fuel cost data in Texas are obtained and part of the data is shown in Table II as an example.

Based on the boxplot of NG shown in Fig. 1, the fuel cost distributions of NG varies month by month. Furthermore, there is a three-month delay for EIA to publish the fuel cost information for the plants. In other words, the most updated fuel cost information can be obtained for estimating the fuel cost distribution in Nov. 2015 is three months ago, i.e., the data of Aug. 2015. As shown in Fig. 1, the distributions between Aug. 2015 and Nov. 2015 have large differences. A more accurate NG distribution estimation is crucial and necessary.

Table II  FUEL COST DATA OF TEXAS

| Date | Plant ID | Plant State | Energy Source | FC' ($/MMBtu) |
|---|---|---|---|---|
| 201301 | 127 | TX | SUB | 2.13 |
| 201302 | 127 | TX | SUB | 2.10 |
| 201303 | 127 | TX | SUB | 2.09 |
| 201304 | 127 | TX | SUB | 2.10 |
| 201305 | 127 | TX | SUB | 2.13 |

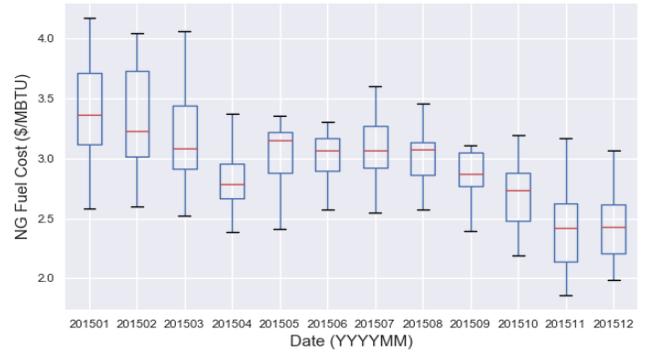

Fig. 1.  Boxplot of NG fuel cost of Texas in 2015.

In contrast, the fuel cost distributions of SUB stays fairly stable, according to Fig. 2. In other words, it is still sufficient and effective to estimate a proper SUB distribution by using a 3-month delayed data. Meanwhile, BIT, as another energy source of coal, has a similar pattern of fuel cost distributions as SUB.

Therefore, the probability density prediction and estimation will focus on NG instead of SUB and BIT in this paper.

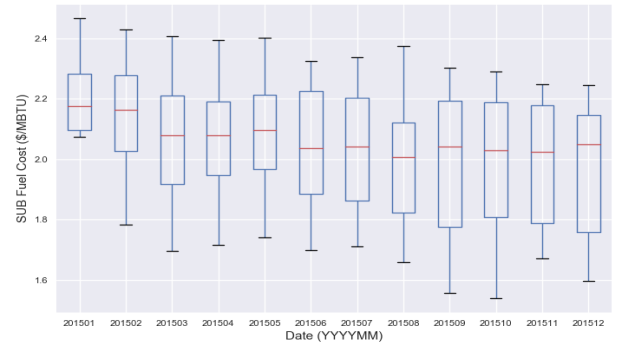

Fig. 2.  Boxplot of SUB fuel cost of Texas in 2015.

## C. Natural Gas Fuel Cost Analysis

According to the previous data exploration and analysis, the primary goal is to study the utility fuel cost of natural gas and estimate the probability density function for the next month in lack of updated information from EIA.

In order to forecast and estimate the NG fuel cost distribution of all 37 NG consuming plants in Texas, the NG monthly fuel costs in each plant are considered as a time series. However, due to 8 of those plants miss more than two months of fuel cost data during 2013-2016, the data from the rest 29 plants are investigated for forecasting. For those plants only miss one data point, the new data is interpolated by (4):

$$FC'_{k,new} = \frac{FC'_{k-1} + FC'_{k+1}}{2}, \quad k = 1,2,\ldots \quad (4)$$

where $FC'_{k,new}$ is the new interpolated fuel cost at time $k$; $FC'_{k-1}$ and $FC'_{k+1}$ are the available fuel cost before and after time $k$, respectively. With the monthly fuel cost series of 29 plants, the ARIMA model is capable of forecasting the fuel cost three-step ahead for each plant.

In addition to the plant's fuel cost information, the more recent electricity used NG spot price at the objective region, which is related to the corresponding plants' fuel cost, is able to be obtained as well. For example, the monthly aggregated NG price of Texas can be found as Henry Hub Natural Gas Spot Price [9].

Both the plant's fuel cost and the aggregated monthly Henry Hub NG price series are used for forecasting the next month fuel cost and estimating the probability distribution. Since the aggregated NG price is able to provide a more recently updated fuel cost data, it is considered to enhance plant's fuel cost prediction by taking the differential between aggregated price at the hub and the plant's fuel cost. The series representing the differential between the aggregated NG price and plants' fuel cost is extracted by (5):

$$\Delta FC'_t = FC'_t - FC^{hub}_t, \quad t = 1,2,\ldots \quad (5)$$

where $\Delta FC'_t$ is the differential value between the $FC'_t$ and $FC^{hub}_t$ at time $t$.

With the differential monthly fuel cost series $\Delta FC'_t$ of 29 plants, the ARIMA model is implemented to forecast the fuel cost three-step ahead. As shown in Fig. 3, the next month differential fuel cost $\Delta FC'_{t+1}$ is required to be forecasted from the three-month delayed fuel cost $\Delta FC'_{t-2}$. Meanwhile, the next month aggregated fuel cost $FC^{hub}_{t+1}$ at the NG hub is also required a one-step ahead prediction. When $\Delta FC'_{(t-2)+3}$ and $FC^{hub}_{t+1}$ are predicted by implementing the ARIMA model, the next month forecasting $FC'_{t+1}$ can be obtained by (6).

$$FC'_{t+1} = \Delta FC'_{(t-2)+3} + FC^{hub}_{t+1}, \quad (6)$$

where $FC^{hub}_{t+1}$ is the one-step ahead forecasting NG fuel cost of Henry Hub at time $t+1$ and the $\Delta FC'_{(t-2)+3}$ is the three-step ahead forecasting differential fuel cost series at time $t+1$.

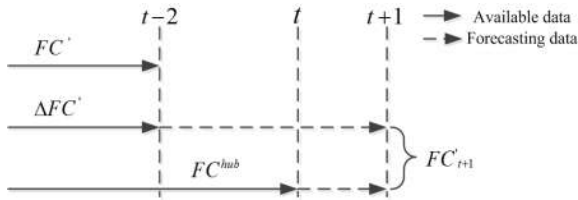

Fig. 3. $FC'$ forecasting process diagram.

The three fuel cost time series $FC'$, $FC^{hub}$ and $\Delta FC'$, will be used in the ARIMA model to forecast the next month fuel cost of each plant. While these forecasted fuel costs of the plants are obtained by each corresponding ARIMA model, the costs will be fitted into Normal Distribution to estimate the next month fuel cost distribution. The numerical results will be given in Section IV.

### III. MODEL DEVELOPMENT

Given the differential fuel cost $\Delta FC'$ and $FC^{hub}$ without any seasonal pattern, a standard ARIMA model is proposed in this paper for forecasting the NG fuel cost series of each plant to estimate the distribution for the next month.

#### A. ARIMA Model

The model used in this paper is an ARIMA model, which is popular for forecasting a stationary time series or a non-stationary time series that can be made to be "stationary" by differencing. The ARIMA model for predicting $\Delta FC'$ and $FC^{hub}$ is a standard $ARIMA(p,d,q)$ model described as follows [10], [11]:

$$\phi_p(B)\nabla^d y_t = \mu + \theta_q(B)\varepsilon_t \quad (7)$$

where $B$ is the backward shift operator, i.e. $B^h y_t = y_{t-h}$; $p$ is the auto-regression (AR) order, which determines how many past values are used for regression; $d$ is the differencing order, which is often used when the stationary assumption is not met; $q$ is the moving-average (MA) order, which determines how many previous error term $\varepsilon_t$ of the process should be considered. The error terms $\varepsilon_t$ are generally assumed to be the independent and identically distributed noise with zero mean and finite variance, i.e., Gaussian white noise; $\mu$ is a constant term.

$$\phi_p(B) = 1 - \phi_1(B) - \phi_2(B^2) - \cdots - \phi_p(B^p) \quad (8)$$

$$\nabla^d = (1-B)^d \quad (9)$$

$$\theta_q(B) = 1 - \theta_1(B) - \theta_2(B^2) - \cdots - \theta_q(B^q) \quad (10)$$

Furthermore, $\phi_1 \ldots \phi_p$ and $\theta_1 \ldots \theta_q$ are the coefficients of the autoregressive and the moving average polynomials, respectively.

#### B. Normal Distribution Fitting

Since the normal distribution in (11) has already been clarified and proven to be a good selection for estimating the fuel cost distribution in [3]. This paper keeps using the standard normal distribution on fitting the data with maximum likelihood method to estimate the probability density function.

$$F(X) = \frac{1}{\sigma\sqrt{2\pi}} e^{-\frac{(x-\mu)^2}{2\sigma^2}} \quad (11)$$

where $\sigma$ and $\mu$ represent the standard deviation and mean of the fitted density function, respectively.

#### C. Kullback-Leibler (KL) Divergence of Normal Distributions

To evaluate the forecasting performance with fitted normal distribution, the symmetric KL divergence [12] between two distributions is formulated (12):

$$D_{p,q} = KL(p||q) + KL(q||p) \quad (12)$$

where $p$ and $q$ are two probability distributions and $D_{p,q}$ stands for the divergence between $p$ and $q$; $KL(p,q)$ for the continuous probability distributions is expressed as (13):

$$KL(p||q) = -\int p(x)\log q(x)dx + \int p(x)\log p(x)dx \quad (13)$$

Since the fitted probability distribution in this paper is normal (Gaussian) distribution, which means $p(x) = N(\mu_1, \sigma_1)$ and $q(x) = N(\mu_2, \sigma_2)$, $KL(p||q)$ can be derived as:

$$KL(p||q) = \log\frac{\sigma_2}{\sigma_1} + \frac{\sigma_1^2 + (\mu_1 - \mu_2)^2}{2\sigma_2^2} - \frac{1}{2} \quad (14)$$

## IV. PREDICTION PERFORMANCE

The proposed prediction model has been applied to forecast the NG fuel cost in Texas. The training data set, from Jan. 2013 to June 2016, including 42 months of fuel cost data and NG hub spot price data are used to obtain the parameters of the ARIMA model. The maximum likelihood estimation method is used to fit the model parameters with the training dataset. In this paper, the toolboxes in Python [13] are used to estimate the model parameters and forecast.

Six months of NG fuel cost data from July 2016 to Dec. 2016 are employed for validating the model. Then the normal distribution is used to fit the data to obtain an estimated distribution of the fuel cost. Meanwhile, the symmetric KL divergence is used to evaluate the distribution forecasting algorithm performance. The forecasting results show the developed algorithm makes more accurate predictions than using the three-month delayed EIA-923 data when the NG market does not have high volatility.

### A. Data Preprocessing

In order to reduce the data fluctuation before fitting the ARIMA model, the natural logarithm transformation for both the Henry hub NG price and plants' fuel cost, given in (15) and (16), are implemented:

$$FC\_log_t^{hub} = \log(FC_t^{hub}) \quad (15)$$

$$\Delta FC\_log_t' = \log(\Delta FC_t' + c) \quad (16)$$

where $c$ is a positive constant offset adding on $\Delta FC_t'$ to guarantee positive value for the logarithm transformation, because there exists negative differential price value between plant's fuel cost and NG hub spot price obtained by (5).

Aiming at forecasting $\Delta FC_{(t-2)+3}'$ of each plant and the NG hub fuel cost $FC_{t+1}^{hub}$, the ARIMA models for these two types time series are described as (17) and (18):

$$\phi_p(B)\nabla^d \Delta FC\_log_t' = \mu + \theta_q(B)\varepsilon_t \quad (17)$$

$$\phi_p(B)\nabla^d FC\_log_t^{hub} = \mu + \theta_q(B)\varepsilon_t \quad (18)$$

By applying Bayesian Information Criterion (BIC) [14] and observing the Autocorrelation Function (ACF) and Partial Autocorrelation Function (PACF) plots [11], the model order is selected to be $ARIMA(2,1,1)$ for predicting the processed Henry hub NG spot price $FC\_log_t^{hub}$. Similarly, either $ARIMA(2,1,1)$ or $ARIMA(2,0,1)$ is selected in order to predict the differential fuel costs $\Delta FC\_log_t'$ of the objective plants in Texas. Since the $\Delta FC\_log_t'$ series of some plants are stationary, the series differencing is not necessary and the $ARIMA(2,0,1)$ is selected for them. Otherwise, $ARIMA(2,1,1)$ is implemented.

### B. Results Comparison

After the $ARIMA(2,1,1)$ is implemented for Henry hub NG spot price prediction, the rolling one-step ahead prediction results from July 2016 to Dec. 2016 are shown in Fig. 4. Compared to the real NG hub spot price, the predictions are very accurate when low volatility presents between July 2016 and Oct. 2016 and get worse when the spot price bounces heavily in Nov. 2016 and Dec. 2016. Since the three-step ahead fuel cost prediction of each plant is based on the obtained NG hub spot price and its one-step ahead prediction, the high volatility and low forecasting accuracy of NG hub spot price are possible to lead a poor distribution estimation performance in these two months.

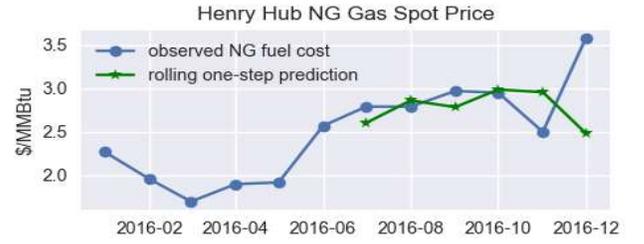

Fig. 4. Time series comparison of NG hub fuel cost

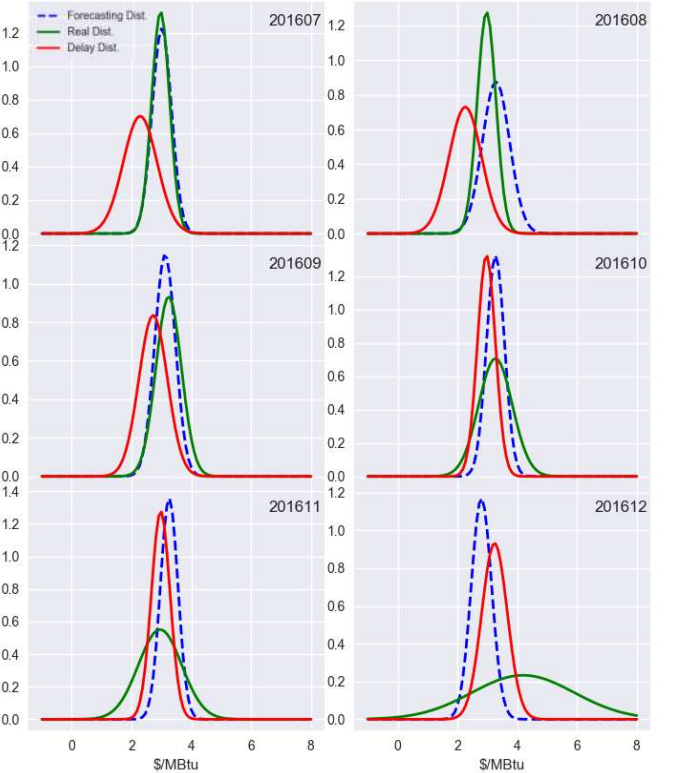

Fig. 5. Fitted Normal Distributions for three-month delay fuel costs, forecasting fuel costs and real fuel costs.

When the prediction results between July 2016 and Dec. 2016 are obtained, they are fitted to normal distributions, as shown in Fig. 5. At the same time, the corresponding fitted three-month delayed distribution and real fuel cost distribution are plotted in Fig. 5 as well.

By observing these distributions between July 2016 and Sept. 2016, the proposed distribution forecasting method using the proposed ARIMA model achieves much better results than the method estimating the distribution by using the plants' fuel cost data from three month ago. In Oct. and Nov., 2016, both the forecasted distribution and the delayed distribution are very close to the real one. However, in Dec. 2016, the forecasting method gives a worse result than the delayed data.

The mean of each fitted normal distribution are presented in Table III. It can be found that the proposed ARIMA forecasting method is capable of reducing the error by about 10 to 20 % when the overall NG market price is steadily increasing from 2.97 $/MMBtu in July 2016 to 3.27 $/MMBtu in Oct. 2016. However, when NG market price unexpectedly dropped to 2.95 $/MMBtu in Nov. 2016 and bounced back to 4.2 $/MMBtu, the forecasting model does not perform well. The reason is that forecasting algorithm turns the prediction price lower for Dec. 2016 to catch up the decreasing price while it receives the price dropping signal from Nov. 2016. Hence, when the overall NG market prices are volatile without any pattern, the forecasting algorithm proposed in this paper does not achieve a better result than just using the delayed data.

Table III  MEAN AND CORRESPONDING ERROR OF FITTED DISTRIBUTIONS

| Month | $Mean_{real}$ | $Mean_{delay}$ | Error | $Mean_{forecast}$ | Error |
|---|---|---|---|---|---|
| 201607 | $2.97 | $2.29 | 22.9% | $3.01 | 1.3% |
| 201608 | $2.98 | $2.26 | 24.2% | $3.29 | 10.4% |
| 201609 | $3.24 | $2.72 | 16.0% | $3.12 | 3.7% |
| 201610 | $3.27 | $2.97 | 9.1% | $3.27 | 0% |
| 201611 | $2.95 | $2.98 | 1% | $3.26 | 10.5% |
| 201612 | $4.20 | $3.24 | 22.9% | $2.8 | 33% |

Table IV  SYMMETRIC KL DIVERGENCE OF ESTIMATED DISTRIBUTIONS

| Month | $D_{real,delay}$ | $D_{real,forecast}$ | Improve |
|---|---|---|---|
| 201607 | 4.22 | 0.03 | 4.19 (99.3%) |
| 201608 | 4.13 | 1.05 | 3.08 (74.6%) |
| 201609 | 1.34 | 0.19 | 1.15 (85.8%) |
| 201610 | 1.54 | 0.90 | 0.64 (41.6%) |
| 201611 | 1.78 | 2.74 | -0.96(-53.9%) |
| 201612 | 9.69 | 20.33 | -10.64(-109.8%) |

Since the data listed in Table III only considers the mean, KL divergence formulated in (15) is implemented to check the divergence for the distributions of three-month delay fuel costs, forecasting fuel cost and the actual fuel cost. The KL divergence numerical results are listed in Table IV. In information theory, the KL divergence is used to measure the difference between two probability distributions over same variable. The higher value of $D_{p,q}$ indicates the larger diversity between distribution $p$ and $q$. In Table IV, the forecasting algorithm is able to improve the estimation by 99.3%, 74.6%, 85.8% and 41.6% while estimation becomes worse in Nov. and Dec. of 2016.

The reason why the three-month delay estimated distributions outperform the forecasting distributions is price bouncing back to the previous level after a suddenly decreasing, which causes an imprecise NG hub spot price prediction. The price data from three months ago was still at a similar level, while the forecasting algorithm determines that is a turning point in Nov. 2016 and provides a declining price prediction in Dec. 2016.

V. CONCLUSION

An ARIMA model was developed in this paper for forecasting the fuel cost and obtaining more accurate distribution estimation. Based on the data exploration, the NG fuel cost in Texas was selected as an example to develop and validate the proposed fuel cost forecast algorithm. The results show the proposed forecasting algorithm has a superior performance over the method that uses the three-month delayed data. Especially when the low volatility presents in the forecasting time domain, the proposed method is capable of achieving a very accurate estimation on the distributions mean with errors of 1.3%, 10.4%, 3.7% and 0%. The results also indicate the good capability of the method in handling the prediction after the turning point. The proposed algorithm is able to be extended and implemented in other states and for other types of fuel costs when there are sufficient data.

REFERENCES

[1] Z. Guo, Q. Wang, M. Fang, Z. Luo, and K. Cen, "Thermodynamic and economic analysis of polygeneration system integrating atmospheric pressure coal pyrolysis technology with circulating fluidized bed power plant," *Applied Energy*, vol. 113, no. Supplement C, pp. 1301 – 1314, 2014.

[2] S. K. Guttikunda and P. Jawahar, "Atmospheric emissions and pollution from the coal-fired thermal power plants in india," *Atmospheric Environment*, vol. 92, pp. 449 – 460, 2014.

[3] C. Fu, C.Wang, C. Miller, and Y. Wang, "Characterization and distribution model for fuel costs in electricity market," 2017, International Energy & Sustainability Conference 2017 (IESC 2017).

[4] E. I. Administration. (2017) Form eia-923 detailed data. Available Online: https://www.eia.gov/electricity/data/eia923/ Access Date: November 2017

[5] Locational Emission Estimation Methodology, available online at: http://www.leem.today/, January 2018

[6] P. Mishura, "Forecasting Natural Gas Price – Time Series and Nonparametric Approach", *World Congress on Engineering*, July 2012

[7] J. Jin and J. Kim, "Forecasting Natural Gas Prices Using Wavelets, Time Series, and Artificial Neural Networks", *PLoS ONE* 10(11): e0142064. doi:10.1371/journal.pone.0142064, 2015

[8] Z. Zhao, C. Wang, M. Nokleby and C. J. Miller, "Improving Shor-Term Electricity Price Forecasting Using Day-Ahead LMP with ARIMA Models", *IEEE PES General Meeting*, July 2017

[9] Henry Hub Natural Gas Spot Price, Available Online: https://www.eia.gov/dnav/ng/hist/rngwhhdD.htm, Access Date: November 2017

[10] G. E. P. Box, G. M. Jenkins and G. C. Reinsel, *Time Series Analysis: Forecasting and Control*, 4th Edition, Published by John Wiley & Sons, Inc., Hoboken, New Jersey, 2008.

[11] R. H. Shumway and D. S. Stoffer, *Time Series Analysis and Its Applications: with R Examples,* Third Edition, Springer New York Dordrecht Heidelberg London, 2011.

[12] D. MacKay, *Information Theory, Inference, and Learning Algorithm*, Version 7.2(fourth printing), Published by Cambridge University Press, 2005

[13] Python Software Foundatoin, Python Language Reference, version 3.6. Available Online: https://www.python.org/ , Access Date: November 2017

[14] The Bayes Information Criterion (BIC), Available Online: http://www-math.mit.edu/~rmd/650/bic.pdf, Access Date: November 2017.